\documentclass[11pt,reqno]{amsart}
\usepackage{graphicx} 
\usepackage{amsmath,setspace}
\usepackage{natbib}                             
\usepackage[colorlinks=true, citecolor=blue]{hyperref}  

\usepackage{natbib}
\usepackage{amssymb}
\usepackage{array,color}
\usepackage{amsfonts,amssymb,amsthm}
\usepackage[margin=3.5cm]{geometry}
\newtheorem{defn}{Definition}[section]
\newtheorem{ex}{Example}[section]

\newtheorem{Pro}{Property}[section]

\newtheorem{thm}{Theorem}[section]
\newtheorem{resu}{Result}[section]
\usepackage{caption}
\usepackage{subcaption}
\usepackage{multirow}
\usepackage{url}
\usepackage{booktabs}
\usepackage{caption}
\usepackage{placeins}
\usepackage[font=small,labelfont=bf]{caption}
\usepackage[labelformat=simple]{subcaption}

\begin{document}
\doublespace

\title[]{On Relative Cumulative Residual Information Measure and Its Applications}
\author[]%
{M\lowercase{ary} A\lowercase{ndrews}$^{\lowercase{a}},$ {S\lowercase{mitha} S.$^{\lowercase{a}}$ \lowercase{and} S\lowercase{udheesh} K. K\lowercase{attumannil}$^{\lowercase{b},}$$^{\dag}$\\
 $^{\lowercase{a}}$K E C\lowercase{ollege} M\lowercase{annanam,} K\lowercase{erala}, I\lowercase{ndia},\\
 $^{\lowercase{b}}$I\lowercase{ndian} S\lowercase{tatistical} I\lowercase{nstitute},
  C\lowercase{hennai}, I\lowercase{ndia.}}}
\thanks{{$^{\dag}$}{Corresponding E-mail: \tt skkattu@isichennai.res.in }}
\maketitle
\vspace{-0.2in}
\begin{abstract}
    We develop a relative cumulative residual information measure (RCRI) that aims to quantify the divergence between two survival functions. The dynamic relative cumulative residual information (DRCRI) measure is also introduced.  We establish some characterization results under the assumption of the proportional hazards model. Additionally,  we obtained the non-parametric estimators of RCRI and DRCRI measures based on the kernel density type estimator for the survival function. The effectiveness of the estimators are assessed through an extensive Monte Carlo simulation study. We consider data from the third Gaia data release (Gaia DR3) to demonstrate the use of the proposed measure. For this study, we have collected epoch photometry data for the objects Gaia DR3 4111834567779557376 and Gaia DR3 5090605830056251776. RCRI-based image analysis is conducted using Chest X-ray data from the publicly available dataset.

\end{abstract}

\section{Introduction}
The concept of entropy was introduced by \citet{shannon1948mathematical} in his seminal work on information theory as a fundamental measure of uncertainty or randomness within a probability distribution. Entropy quantifies the average amount of information produced by a random variable. Shannon's entropy has found extensive applications in signal processing, image processing, reliability engineering, medical image analysis, risk theory, economics etc. The Shannon entropy measure associated with a nonnegative random variable $X$ is defined as
$$H(X)=-\int_{0}^{\infty}f(x)\log f (x)dx, $$ where `log' denotes the natural logarithm.


Different measures discuss different aspects of entropy. Several divergence measures are introduced in the literature to study the behavior of two random variables as a natural extension of entropy. Let $X$ and $Y$ be two non-negative random variables having probability density  functions $f(x)$ and $g(x)$, respectively. \citet{kullback1951information} have extensively studied the concept of directed divergence which aims at discrimination between two populations and is given by
\begin{align*}
    D(f||g)=\int_{0}^{\infty} f(x) log\left(\frac{f(x)}{g(x)}\right) \, dx.
\end{align*}
For some recent works in this area, one can refer to \citet{zohrevand2020adjusted}, \citet{mehrali2021parameter}.\\
According to the work of \citet{lad2015extropy}, while entropy quantifies the disorder or uncertainty within a system, it may not capture all the characteristics of a distribution. To complement this, the concept of extropy was introduced, offering additional descriptive dimensions. It is formally defined as
\[
J(X) = -\frac{1}{2} \int_{0}^{\infty} f^2(x) \, dx.
\]

Weighted extropy was introduced and explored by \citet{balakrishnan2022weighted}. In addition, \citet{jahanshahi2020cumulative} proposed the concept of cumulative residual extropy (CRJ) and examined its theoretical foundations. For a random variable \(X\), CRJ is defined as
\[
\text{CRJ}(X) = -\frac{1}{2} \int_{0}^{\infty} \bar{F}^2(x) \, dx.
\]

The moment generating function (MGF) is widely regarded as a highly effective tool for deriving the moments of a probability distribution. Building upon this well-established idea, \citet{golomb1966information} proposed the information generating function (IGF) for a distribution, expressed as
\[
\text{IG}_\alpha(X) = \int_{0}^{\infty} f^\alpha(x) \, dx.
\]

Later, \citet{guiasu1985relative} made a notable advancement by formulating the relative information generating function (RIGF). This essential concept applies to two continuous random variables \(X\) and \(Y\), characterized by their respective probability density functions \(f(\cdot)\) and \(g(\cdot)\). The RIGF of \(X\) relative to the reference variable \(Y\) is defined as
\[
\text{RIG}_\alpha(X, Y) = \int_{0}^{\infty} f^\alpha(x) g^{1-\alpha}(x) \, dx.
\]
Another useful measure for discrimination among distributions is the notion of Chernoff distance, which finds application in several branches of learning as a potential measure of distance between two populations.
\begin{align*}
    C(f, g) = -\log \int f^\alpha(x) g^{1-\alpha}(x) \, dx, \quad 0 < \alpha < 1.
\end{align*} \citet{asadi2005dynamic} have studied the application of this measure in the context of reliability studies. \citet{ghosh2018chernoff} and \citet{kayal2018quantile} have also made significant contributions to this area.
Subsequently, the concept of the cumulative residual information generating function (CRIGF) was introduced by \citet{chakraborty2024weighted}, which is given by
\[
\text{CRIG}_\alpha(X) = \int_{0}^{\infty} \bar{F}^\alpha(x) \, dx.
\]

\par The distribution function is more regular than the density function since it is defined in an integral form, whereas the density function is computed using the derivative of the distribution function.
There are certain limitations to using Shannon's entropy to measure randomness in some systems. Alternative entropy measures, such as cumulative residual entropy \citet{rao2004cumulative} and cumulative entropy \citet{di2009cumulative}, are more suited for specific applications, such as lifetime analysis. Additionally, weighted versions of these measures were developed by \citet{mirali2017weighted}  to address different contexts.  For a non-negative random variable $X$ with distribution function $F(x)$, the cumulative residual entropy, which quantifies the uncertainty about the remaining lifetime of a system, is defined as follows.
\begin{align*}
\xi = -\int_0 ^ {\infty} \Bar{F}(x) \log \Bar{F}(x) dx.
\end{align*}
See \citet{kattumannil2022generalized} and the references therein, for the recent development in this area. \citet{park2012cumulative}  defined cumulative Kullback–Leibler information, which can be viewed as the analog of the Kullback–Leibler information concerning the cumulative distribution function and is given by
\begin{align*}
    CRKL(G:F)=\int_0 ^ {\infty} \Bar{G}(x) \log\left(\frac{\Bar{G}(x)}{\Bar{F}(x)}\right) dx - (E(Y)-E(X)).
\end{align*} See \citet{baratpour2012testing} for the properties of $CRKL(G:F)$.

In survival analysis and life testing, considering the current age of the system is very important. So when assessing uncertainty or distinguishing between systems, traditional measures like Shannon's entropy and other distance and divergence measures may not be appropriate. In such cases, a more realistic approach for measuring the uncertainty is to define divergence measures about the remaining lifetime of the unit. This was studied thoroughly by \citet{ebrahimi1995new}. For some  developments in this area, one can refer to \citet{cali2017some}, \citet{kharazmi2021cumulative}, and the references therein.

\par Several works were done using cumulative and dynamic cumulative residual information generating measures. \citet{kharazmi2021cumulative} introduced the cumulative residual entropy generating function and explored its relationship with the Gini mean difference. \citet{capaldo2024cumulative} introduced and studied the cumulative information generating function, which provides a unifying mathematical tool suitable to deal with classical and fractional entropies based on the cumulative distribution function and on the survival function. \citet{smitha2024dynamic} have done an extensive study regarding the dynamic cumulative residual entropy generating function (DCREGF) and proposed some characterization results using the relationship between DCREGF and basic reliability concepts. They also proposed a new class of life distributions based on decreasing DCREGF, developed a test for decreasing DCREGF, and studied its performance. \citet{smitha2026weighted} defined the weighted cumulative residual entropy generating function (WCREGF) and studied its properties. They also introduced the dynamic weighted cumulative residual entropy generating function (DWCREGF). However,  few studies were carried out in the area of relative cumulative information generating function.  Motivated by this, in the present paper we introduced and studied the properties of the relative cumulative residual information (RCRI) measure and its dynamic version.

 The rest of the paper is structured as follows. In Section 2, the relative cumulative residual information (RCRI) measure is introduced, while Section 3 discusses the dynamic relative cumulative residual information (DRCRI) measure. We also discuss the characterization results based on DRCRI. Section 4 addresses the non-parametric kernel estimation of RCRI and DRCRI measure. In Section 5, we carry out Monte Carlo simulation studies to assess the finite sample performance of the proposed estimators. Section 6 presents the analysis of  real-life data, where we consider astronomical data from the ESA (European Space Agency) Gaia mission. Epoch photometry data of two objects (Gaia DR3 4111834567779557376 and Gaia DR3 5090605830056251776) were used for this purpose. Section 7 delve into the image processing application of RCRI measure. The concluding remarks are given in Section 8.

\section{Relative Cumulative Residual Information measure}
We discuss the concept of information generating measure concerning two random variables, namely relative cumulative residual information (RCRI) measure, and then study its properties. Next, we define RCRI measure.
\begin{defn}
 Let  $X$ and $Y$ be two non-negative random variables having survival functions $\bar{F}(x)$ and $\Bar{G}(x)$ respectively. Then the relative cumulative residual information measure between $X$ and $Y$   is defined as
\begin{equation}
    R_{\alpha, \beta}(\bar{F},\bar{G}) = \int_0^\infty (\bar{F}(x))^\alpha (\bar{G}(x))^\beta \, dx, \quad \alpha, \beta > 0.
\end{equation}
\end{defn}
From the definition of RCRI measure, we can see that it is a divergence measure. That is, in general, RCRI measure is not symmetric and is symmetric when $\alpha =\beta$. This fact is illustrated by considering two exponential distributions.
\vspace{-0.1in}
\begin{ex}
   Let $X$ and $Y$ be two exponential random variables with parameters $\lambda_1$ and $\lambda_2$ respectively, then the  RCRI measure between $X$ and $Y$ becomes
\begin{align*}
R_{\alpha,\beta}(\Bar{F}, \Bar{G}) =\frac{1}{\alpha\lambda_1 + \beta \lambda_2} .
\end{align*}
However
\begin{align*}
R_{\alpha,\beta}(\Bar{G}, \Bar{F})= \frac{1}{\alpha\lambda_2+\beta\lambda_1}.
\end{align*}
Therefore,  $R_{\alpha,\beta}(\Bar{F}, \Bar{G})  \neq R_{\alpha,\beta}(\Bar{G}, \Bar{F}) $.
 \end{ex}
\noindent Next, we study the properties of RCRI measure.
\begin{Pro}
 When $ \Bar{F} (x)=\Bar{G}(x)$,
the proposed measure becomes
\begin{equation}\label{CREGF}
    R_{\alpha, \beta}(\Bar{F})=\int_0^\infty \left(\bar{F}(x)\right)^{\alpha +\beta}dx,
\end{equation}
which is the cumulative residual entropy generating function introduced by \citet{kharazmi2021cumulative}.
\end{Pro}
\noindent  See \citet{smitha2024dynamic} for more details on $R_{\alpha, \beta}(\Bar{F})$.
   Next, using the arithmetic mean and geometric mean inequality, we obtain an upper bound for  RCRI measure  in terms of cumulative residual entropy generating functions.
\begin{Pro}
    Suppose that $X$ and $Y$ are two non-negative random variables having finite means, then
        \begin{align*}
       R_{\alpha, \beta}(\Bar{F}, \Bar{G}) \leq \left( R_{2\alpha}(\bar{F}) + R_{2\beta}(\bar{G}) \right)
.    \end{align*}
    \textbf {Proof}: Using arithmetic mean and geometric mean inequality
     \[
\int_0^\infty (\bar{F}(x))^\alpha (\bar{G}(x))^\beta \, dx
\leq \frac{1}{4} \int_0^\infty \left( (\bar{F}(x))^\alpha + (\bar{G}(x))^\beta \right)^2 dx
\]
\[
= \frac{1}{4} \left( \int_0^\infty (\bar{F}(x))^{2\alpha} dx + \int_0^\infty (\bar{G}(x))^{2\beta} dx + 2 \int_0^\infty (\bar{F}(x))^\alpha (\bar{G}(x))^\beta dx \right).
\]Hence
\[
\frac{1}{2}\left(\int_0^\infty (\bar{F}(x))^\alpha (\bar{G}(x))^\beta dx \right)\leq \frac{1}{4} \left( \int_0^\infty (\bar{F}(x))^{2\alpha} dx + \int_0^\infty (\bar{G}(x))^{2\beta} dx \right)
\]
\[
\int_0^\infty (\bar{F}(x))^\alpha (\bar{G}(x))^\beta dx \leq \frac{1}{2} \left( \int_0^\infty (\bar{F}(x))^{2\alpha} dx + \int_0^\infty (\bar{G}(x))^{2\beta} dx \right)
\]
\[R_{\alpha, \beta}(\Bar{F}, \Bar{G}) \leq \left( R_{2\alpha}(\bar{F}) + R_{2\beta}(\bar{G}) \right)\]
where $R_{2\alpha}(\bar{F})$ and $R_{2\alpha}(\bar{G})$ are the cumulative entropy generation functions.
\end{Pro}
In the following theorem, we gave an approximation for RCRI measure in terms of cumulative residual entropy generating function.
 \begin{thm}
 Let $X$ be a non-negative random variable with survival function $\Bar{F}(x;\theta)$ and probability density function $f(x;\theta)$, which is differentiable at $\theta$. Let $K$ be a real constant, $\Delta>0$,  and $\Delta\theta$ be very small, then
       \begin{equation}
           R_{\alpha,\beta}(\Bar{F}(x,\theta) , \Bar{F}(x;\theta+\Delta\theta) \simeq R_{\alpha,\beta}(\Bar{F})+K.\Delta\theta, \\
       \end{equation}
       where $R_{\alpha,\beta}(\Bar{F})$ is the cumulative residual entropy generating function given in (\ref{CREGF})
       \end{thm}
      \textbf {Proof}: Using Taylor Series expansion
\begin{eqnarray*}
    R_{\alpha,\beta}(\overline{F}(x,\theta), \overline{F}(x;\theta+\Delta\theta)) &&= \int_0^\infty \left( \overline{F}(x,\theta) \right)^\alpha \left( \overline{F}(x,\theta+\Delta\theta) \right)^\beta \, dx
\\
&&= \int_0^\infty \left( \overline{F}(x,\theta) \right)^\alpha \left( \overline{F}(x,\theta) +\frac{\Delta\theta}{1!}(-f(x,\theta)) + \ldots \right)^\beta \, dx \\
&&\approx \int_0^\infty \left( \overline{F}(x,\theta) \right)^\alpha \left( \overline{F}(x,\theta)- \Delta\theta f(x,\theta)\right)^\beta \, dx,
\end{eqnarray*}
where higher order powers of \(\Delta\theta\) are neglected.
\begin{eqnarray*}
    R_{\alpha,\beta}(\overline{F}(x,\theta), \overline{F}(x;\theta+\Delta\theta)) &&\approx\int_0^\infty \left( \overline{F}(x,\theta) \right)^\alpha \Big( \left( \overline{F}(x,\theta) \right)^\beta
\\&& \qquad- \beta c_1 \left( \overline{F}(x,\theta) \right)^{\beta-1} \frac{\Delta \theta}{1!}
f(x,\theta)+  \dots \Big) \, dx\\
&&\approx \int_0^\infty \left( \overline{F}(x,\theta) \right)^{\alpha} \left( \overline{F}(x,\theta) \right)^\beta dx-\\&&\qquad \beta c_1 \frac{\Delta\theta}{1!} \int_0^\infty \left( \overline{F}(x,\theta) \right)^{\alpha+\beta-1}f(x,\theta)) dx \\
&&\approx \int_0^\infty \left( \overline{F}(x,\theta) \right)^{\alpha+\beta}dx+\beta c_1 \Delta\theta \int_0^1 u^{\alpha+\beta-1} du \\
&&\approx \int_0^{\infty} \left( \overline{F}(x,\theta) \right)^{\alpha+\beta}dx +
\frac{\beta}{\alpha+\beta} \Delta \theta.
\end{eqnarray*}That is,
\begin{equation*}
 R_{\alpha,\beta}(\Bar{F}(x,\theta) , \Bar{F}(x;\theta+\Delta\theta) \simeq R_{\alpha,\beta}(\Bar{F}) + K.\Delta\theta,
 \end{equation*}
where
  $K=\frac{\beta}{\alpha +\beta}.$
 \\ Under the  PH model assumption, the survival functions of the random variables  $X$ and $Y$ satisfy the relationship given by
 \begin{equation}\label{phm}
 \bar{G}(x)=(\bar{F}(x))^{\theta}; \,\,\theta >0.
 \end{equation}
 We can easily verify that the hazard rate of $Y$ is proportional to that of $X$. That is,\\
\begin{equation*}
h_2(x)=\theta h_1(x),
\end{equation*}
where
\begin{equation*}
  h_1(x) = \frac{f(x)}{\bar{F}(x)} \quad \text{and} \quad h_2(x) = \frac{g(x)}{\bar{G}(x)}.
   \end{equation*}
   We exploit the assumption given in (\ref{phm}) to establish some results given in the subsequent sections.
The RCRI measure under PH model  becomes
   \begin{equation}\label{Rstar}
       R_{\alpha, \beta}(\Bar{F}) =\int_0^\infty \left (\bar{F}(x)\right)^{\alpha +\beta\theta}dx.
   \end{equation}
In Table 1, we presented RCRI measures under PH model assumption for some well-known distributions.

\begin{table}[ht]
\centering
\caption{RCRI measure under PH model assumption.}
\begin{tabular}{ccc}
\toprule
Distribution & Survival Function & RCRI measure  \\
\midrule
Uniform & $(1 - \frac{x}{a})$  , $0<x<a$  & $\frac{a}{\alpha + \beta \theta+1}$ \\
\addlinespace[0.75em] 
Exponential & $e^{-\lambda x}$, $x \geq 0$ , $ \lambda >0$ & $\frac{1}{\lambda (\alpha + \beta \theta)}$ \\
\addlinespace[0.75em] 
Weibull & $e^{-(\lambda x)^k}$ , $x \geq 0$, $ \lambda >0$, $ k >0$ & $\frac{1}{\lambda k} \left( \frac{\Gamma\left(\frac{1}{k}\right)}{(\alpha + \beta \theta)^{\frac{1}{k}}} \right)$ \\
\addlinespace[0.75em] 
GPD & $\left(1 + \frac{ax}{b}\right)^{-(1 + \frac{1}{a})}$ , $x \geq 0$, $a>-1$, $b>0$ & $\frac{b}{ \left(a + 1\right) (\alpha + \beta \theta) - a}$ \\
\addlinespace[0.75em] 
Pareto I & $\left(\frac{k}{x}\right)^a$, $x \geq k $, $a>0$  & $\frac{k}{a (\alpha + \beta \theta) - 1}$ \\
\addlinespace[0.75em] 
Pareto II & $\left(1 + \frac{x}{a}\right)^{-b}$, $x \geq 0$ , $a>0$, $b>0$& $\frac{a}{b (\alpha + \beta \theta) - 1}$ \\
\bottomrule
\end{tabular}
\label{tab:rcri_results}
\end{table}
The next property shows that RCRI measure is shift independent under PH model assumption.

\begin{Pro}
  Let $X$ be a continuous non-negative random variable and $Y=aX+b$, with $a>0$ and $b\geq 0$, then
  $R_{\alpha, \beta \theta}(Y)= a  R_{\alpha, \beta \theta}(X)$.
  \end{Pro}
  \noindent
 This property follows by using the result $ \bar{F}_{aX+b}(x)= \bar{F}_X(\frac{x-b}{a})$ for all $x>b$. \\Next, we discuss the dynamic version of the RCRI measure.
    \section{ Dynamic Relative Cumulative Residual Information (DRCRI) Measure}
    In many practical situations, complete lifetime data are unavailable, and analysis must be based on truncated observations that describe the residual lifetime beyond a specified time point. In such cases, information measures depend on time and are therefore called dynamic information measures. For example, in insurance applications, one may be interested in modelling the remaining lifetime of individuals after a certain age, such as the retirement age.  Many researchers have extended the information measures to the truncated situation \citet{ebrahimi1996characterisation}, \citet{nair2007characterization}. Recent developments in dynamic version of information measures can be found in  \citet{hashempour2024new}, \citet{mohammadi2024dynamic}, \citet{saranya2026relative}, \citet{pandey2026study} and the references therein.  Motivated by this, we define the RCRI measure for truncated random variables. \\
    \begin{defn}
    Let $X$ and $Y$ be two non-negative random variables with survival functions $\Bar{F}(x)$ and $\Bar{G}(x)$ respectively.
    Suppose $X_t={X-t}{\mid X>t}$ and $Y_t={Y-t}{\mid Y>t}$ are the residual random variables corresponding to $X$ and $Y$ respectively. Then the  relative cumulative residual information measure between $X_t$ and $Y_t$
 is defined as
   \begin{equation}\label{R(t)}
       R_{\alpha, \beta}( \Bar{F},\Bar{G},t)=\int_t^\infty \left( \frac{\Bar{F}(x)}{\Bar{F}(t)}\right)^\alpha \left( \frac{\Bar{G}(x)}{\Bar{G}(t)}\right)^\beta  dx,\quad \alpha, \beta>0.
   \end{equation}
   \end{defn}
   Next we study the properties of DRCRI measure. The following result shows the relationship between the dynamic relative cumulative residual information measure and hazard rates.
 \begin{resu}Let $h_1(t)$ and $h_2(t)$  be the hazard rates of $X$  and $Y$ respectively, then we have
\begin{equation}\label{R'}
R_{\alpha,\beta}'(\bar{F},\bar{G},t)= (\beta h_2(t) + \alpha h_1(t)) R_{\alpha,\beta}(\bar{F},{\bar{G}},t)-1.
\end{equation}
\noindent\textbf {Proof}:
\begin{equation*}
    R_{\alpha,\beta}'(\bar{F},\bar{G},t)= \frac{d}{dt}\left(\frac{1}{\bar F(t)^{\alpha}\bar G(t)^{\beta}} \int_t^\infty \bar{F}(x)^{\alpha} \bar{G}(x)^{\beta} dx\right)
\end{equation*}
\begin{equation*}
R_{\alpha,\beta}'(\bar{F},\bar{G},t)= \int_t^\infty \bar{F}(x)^{\alpha} \bar{G}(x)^{\beta} dx \left(-\alpha \frac{f(t)}{\bar F(t)\bar F(t)^{\alpha} \bar G(t)^{\beta}}-\beta \frac{g(t)}{\bar G(t)\bar G(t)^{\beta} \bar F(t)^{\alpha}}\right)-1
\end{equation*}
\begin{align*}
=(\beta h_2(t) + \alpha h_1(t)) R_{\alpha,\beta}(\bar{F},{\bar{G}},t)-1,
\end{align*}
\end{resu}\noindent where prime denotes the derivative of $R_{\alpha,\beta}(\bar{F},\bar{G},t)$ with respect to $t$.
 \begin{resu}
    Under the proportional hazards model specified in (\ref{phm}), we have the relationship between the dynamic relative cumulative residual information measure and hazard rates given by \\
    \begin{equation}\label{ph}
    R_{\alpha,\beta}'(\bar{F},t)= (\beta \theta+\alpha) h_1(t) R_{\alpha,\beta}(\bar{F},t)-1.
    \end{equation}
    \end{resu}
     Next,  we look into the problem of characterizing probability distributions using the functional form of $ R_{\alpha,\beta}(\Bar{F},\Bar{G},t)$.
    First we examine the situation where $R_{\alpha,\beta}(\bar{F},\bar{G},t)$ is independent of $t$.
\begin{thm}
   Let $F(x)$ and $G(x)$ be  absolutely continuous distribution functions and  $R_{\alpha,\beta}(\bar{F},\bar{G},t)$ be as defined in (\ref{R(t)}). If $R_{\alpha,\beta}(\bar{F},\bar{G},t)$ is a positive constant, then $F(x)$ is exponential if and only if $G(x)$ is exponential.
\end{thm}
    \noindent\textbf {Proof}:
      Let $R_{\alpha,\beta}(\bar{F},\bar{G},t)=c$, where $c$ is a positive constant and that $F(x)$ is the exponential distribution with  survival function
      $$\Bar{F}(x)= e^{-\lambda x},\, x>0,\,\lambda>0.$$
      By using the relationship between  $R_{\alpha,\beta}(\bar{F},\bar{G},t)$ and hazard rates, we obtain
     $$c(\beta h_2(t) +\alpha \lambda)=1.$$
     The solution to the above equation is
         $$h_2(t)= \frac{\frac{1}{c} -\alpha\lambda}{\beta}=k,\,\frac{1}{c}> \alpha \lambda ,$$  where $k$ is a positive constant.   Hence $G(x)$ is exponential.

         Conversely, assume that
         \begin{align*}
\bar{G}(x) &= e^{-kx}, \, x > 0, k > 0
\end{align*}
and using the relationship given in (\ref{R'}), we have
$$h_1(t) = \frac{1 - kc\beta}{c\alpha}, k\beta<\frac{1}{c}.$$
    Now
    $$\Bar{F}(x) = exp\left({-\int_0^x h_1(t) \, dt}\right) ,$$
   and simplifying we get,
    $$\Bar{F}(x) = exp\left({-\frac{(1 - kc\beta)}{c\alpha} x}\right).$$
Hence $F(x)$ is exponential.

The following theorem focuses on the situation where $R_{\alpha,\beta}(\bar{F},t)$ is a linear function of $t$.
\begin{thm}Let $F(x)$ and $G(x)$ be absolutely continuous distribution functions and $h_1(t)$ be the hazard rate of $X$. Assume that $\left (Y,\Bar{G}\right)$ is the PH model of $\left (X,\Bar{F}\right)$ then $R_{\alpha,\beta}(\Bar{F}, t)$ is a linear function in $t$ if and only if $F(x)$ is generalized Pareto distribution (GPD) with survival function
     \begin{equation}\label{gpd}
         \Bar{F}(x)= \left(1+\frac{b}{a}x\right)^{-\left(1+\frac{1}{b}\right)},  \quad  x>0,\,b>-1, a>0 .
           \end{equation}
           \end{thm}
          \noindent\textbf {Proof}:
          Under the conditions of the theorem, when $X$ has GPD, using (\ref{gpd}) we obtain
     \begin{align*}
R_{\alpha,\beta}(\bar{F}, t) &= \frac{b(a + bt)}{a^2 \left(\left(1 + \frac{1}{b}\right)(\alpha + \beta\theta) - 1\right)} \\
                   &= k(a + bt),
                   \end{align*}
                   where, $ k=\frac{b}{a^2 \left(\left(1 + \frac{1}{b}\right)(\alpha + \beta\theta) - 1\right)}. $
This gives that
$R_{\alpha,\beta}(\bar{F}, t)$ is a linear function in \(t\).\\
Conversely, assume that
\begin{align*}
R_{\alpha, \beta}(\bar{F}, t) = a + bt.
\end{align*}
Differentiating above equation with respect to $t$, we obtain
\begin{align*}
R'_{\alpha, \beta}(\bar{F}, t) = b.
\end{align*}
Under PH model assumption, substituting above two equations in (\ref{ph}), we obtain
\begin{align*}
    b &= (\alpha + \beta \theta) h_1(t) (a + bt) - 1.
   \end{align*}Rearranging, we have
   $$ (a + bt) h_1(t) = \frac{b + 1}{\alpha + \beta \theta}.$$
Differentiating above equation with respect to $t$, we obtain
\begin{equation*}
    (a + bt)h_1'(t)+h_1(t)b = 0.
\end{equation*}
From above, we have
$$\frac{-h_1'(t)}{h _1(t)}=\frac{b}{a+bt}=\frac{1}{k+t}, $$ where $ k=\frac{a}{b}$.\\
We can rewrite the above equation as
     $$\frac{-d}{dt} (\log h_1(t))= \frac{1}{k+t}.$$
Integrating with respect to $t$, we have
\begin{align*}
    -\log h_1(t)=\log(k+t)+\log c.
\end{align*}Or
\begin{equation}\label{eq12}
    h_1(t)=\frac{1}{(k+t)c}=\frac{1}{ct+d},
\end{equation}where $ d=kc$.  \citet{hall1981mean} showed that  (\ref{eq12}) is the characteristic property of the GPD. Thus the necessary part of the theorem is proved.

 In the next theorem, we give a characterization result for GPD based on the relationship between DRCRI measure and hazard rate.
\begin{thm}Under the conditions of Theorem 3.2, the relationship
\begin{equation}\label{relation}
R_{\alpha,\beta}(\bar{F}, t) = k \left(h_1(t) \right)^{-1},
\end{equation}
where $k$ is a positive constant and $h_1(t)$ is the hazard rate of $X$,  holds if and only if  $X$ has GPD with survival function given in (\ref{gpd}).
\end{thm}
\noindent\textbf {Proof}:
Assume that (\ref{relation}) holds and is differentiable with respect to $t.$ Then we have
\begin{align*}
    R_{\alpha,\beta}'(\bar{F}, t)=-k (h_1(t))^{-2} h_1'(t).
\end{align*}Or
\begin{equation}\label{eq14}
     R_{\alpha,\beta}'(\bar{F}, t)= -k\frac{ h_1'(t)}{(h_1(t))^2}.
\end{equation}
Substituting  (\ref{eq14}) in (\ref{ph}), we obtain
\begin{align*}
  (\alpha+\beta\theta)h_1(t) R_{\alpha,\beta}(\Bar{F},t)-1 =k\frac{d}{dt} \left(\frac{1}{h_1(t)}\right).
  \end{align*}
  Hence using (\ref{relation}) we have
  $$\frac{d}{dt} \left(\frac{1}{h_1(t)}\right)=\frac{k(\alpha+\beta\theta)-1}{k}.$$
  Integrating both sides of the above equation  with respect to $t$ we have
   $$ \frac{1}{h_1(t)} = \left( \frac{k(\alpha + \beta \theta) - 1}{k} \right)t + B = At + B$$
where $A$ and $B$ positive constants. Hence, we have
\begin{equation}\label{eq15}
     h_1(t)=\frac{1}{At+B}.
\end{equation}
\citet{hall1981mean} showed that (\ref{eq15}) is the characteristic property of GPD.

Conversely, assume that $X\sim GPD$, by direct calculation we
obtain  \begin{align*}
R(\bar{F}, t) &= \frac{b(a + bt)}{a^2 \left(\left(1 + \frac{1}{b}\right)(\alpha + \beta\theta) - 1\right)} \\
                   &= (\frac{a+bt}{b+1})\frac{(b+1)b}{a^2 \left(\left(1 + \frac{1}{b}\right)(\alpha + \beta\theta) - 1\right)}\\
                   &= k\frac{1}{h_1(t)},
\end{align*}
 where $k = \frac{(b+1)b}{a^2 \left(\left(1 + \frac{1}{b}\right)(\alpha + \beta\theta) - 1\right)}$. Hence we have the proof of the theorem.\\
The next theorem focuses on a characterization result for the GPD  by the form of $R_{\alpha,\beta}(\bar{F},t)$ in terms of the mean residual life function.
    \begin{thm}
Let $X$ be a non-negative random variable, admitting an absolutely continuous distribution function $F$ and with mean residual life (mrl) function  $m_1(t)=E(X-t|X>t)$ and let $G$ be the proportional hazards model of $F$ specified in (\ref{phm}). Then the relationship
\begin{equation}
    R_{\alpha,\beta}(\bar{F}, t) = k~m_1(t), t>0,
\end{equation}
 holds if and only if $X\sim GPD$.
 \end{thm}
\noindent\textbf {Proof}: Assume that
$$R_{\alpha,\beta}(\bar{F}, t) =k~m_1(t).$$ Differentiate both sides of the above equation with respect to $t$, we get
$$R_{\alpha,\beta}'(\bar{F}, t) =k~m_1'(t).$$
Using the relationship between $R_{\alpha,\beta}(\Bar{F},t)$ and hazard rate under the proportional hazards model assumption, given in (\ref{phm}) the above equation becomes
\begin{align*}
    (\alpha+\beta\theta)h_1(t) R_{\alpha,\beta}(\bar{F}, t)-1=k~m_1'(t).
   \end{align*}Or
   \begin{equation}
       \label{eq16}
       (\alpha+\beta\theta)h_1(t)k~m_1(t)-1=k~m_1'(t).
   \end{equation}
We have the relationships between the hazard rate and the mean residual life given by
\begin{equation}\label{eq17}
\frac{1+m_1'(t)}{m_1(t)}=h_1(t).
\end{equation}Combining (\ref{eq16}) and (\ref{eq17}) we obtain
\begin{align*}
    m_1'(t)=\frac{1-k(\alpha+\beta\theta)}{((\alpha+\beta\theta)k-k)}=a,
\end{align*}where $a$ is real constant. This implies that $m_1(t)$ is linear in $t$. Linear mrl function characterises the GPD \citet{hall1981mean}.

Conversely, assume that $X$ follows GPD. By direct calculation,
    $$ R_{\alpha,\beta}(\bar{F},t) =k~m_1(t),$$
 where $ k=\frac{b}{a^2 \left(\left(1 + \frac{1}{b}\right)(\alpha + \beta\theta) - 1\right)}.$  Hence the proof of the theorem.
\section {Non-parametric Kernel Estimation}
 Let $X_1, X_2,\ldots, X_n$ be a random sample from  $F$ and $Y_1,Y_2,\ldots, Y_n$ be a random sample from $G$. Here we find non-parametric estimators for the proposed measures using the kernel density estimator.
 We assume that kernel function $k(x)$ satisfies the following conditions:

 1) $k(x)\geq 0$, for   all $x$

 2) $\int k(x)dx =1 $

 3) $ k(.)$ is symmetric.

 The kernel density estimator of the probability density function $f(x)$ at a point $x$ is given by \citet{parzen1962estimation}
 \begin{align}
    f_n(x) = \frac{1}{nh} \sum_{j=1}^{n} k\left(\frac{x - X_j}{h}\right),
 \end{align}
 where $h$ is the bandwidth.

As our measure is defined using survival functions,  we consider the kernel type estimator of survival function and it is given by
     \begin{align*}
    \bar{F}(x) = \frac{1}{n} \sum_{j=1}^{n} \bar{K} \left( \frac{x - X_j}{h} \right),
\end{align*}
     where $\bar{K}$ denotes the survival function of the kernel $k$, ie. $\bar{K}(t)=\int_t^\infty k(u)du$.

 The non-parametric kernel estimator of
 $RCRI$ measure, $R_{\alpha, \beta}(\Bar{F},\Bar{G})$, can be defined as
 \begin{equation}\label{est1}
     \widehat{R}_{\alpha, \beta}(\widehat{\bar{F}},\widehat{\bar{G}}) = \int_0^\infty \left( \frac{1}{n}\sum_{j=1}^{n} \bar{K}\left( \frac{x - X_j}{h} \right) \right)^\alpha  \left( \frac{1}{n}\sum_{j=1}^{n} \bar{K}\left( \frac{x - Y_j}{h} \right) \right)^\beta dx.
 \end{equation}

 The estimator for DRCRI measure is given as
 \begin{equation}\label{est2}
\widehat{R}_{\alpha,\beta}(\widehat{\bar{F}},\widehat{\bar{G}},t) = \int_t^\infty \left( \frac{ \sum_{j=1}^{n} \bar{K}\left( \frac{x - X_j}{h} \right)}{ \sum_{j=1}^{n} \bar{K}\left( \frac{t - X_j}{h} \right)} \right)^\alpha \left( \frac{ \sum_{j=1}^{n} \bar{K}\left( \frac{x - Y_j}{h} \right)}{ \sum_{j=1}^{n} \bar{K}\left( \frac{t - Y_j}{h} \right)} \right)^\beta \ dx.
 \end{equation}


Next, we study the consistency of the proposed estimators. \citet{berg2009cdf} establish the consistency of the kernel type estimator of cumulative distribution function $F(x)$, where the estimator is given by
\begin{align*}
    \hat{F}_h(x) = \int_{-\infty}^t \hat{f}(t) \, dx = \frac{1}{n} \sum_{j=1}^{n} \tilde{K} \left( \frac{t - X_j}{h} \right).
\end{align*}Here $\tilde{K}(t)=\int_{0}^{t}k(u)du$. \\
For establishing the consistency, \citet{berg2009cdf} has stated the variance of $ \hat F_h(t)$ as
\begin{equation}\label{varbp}
    \operatorname{Var}(\hat{F}_h(t)) = \frac{F(t)(1 - F(t))}{n} - \frac{2 f(t)}{n} \left( \int u \, \tilde{K}(u) k(u) \, du \right) h + O\left(\frac{h}{n}\right).
\end{equation}
Under some assumptions if $h\to 0$ as $n\to \infty$ and $nh\to\infty$, then $\operatorname{Var}(\hat{F}_h(t))$ tends to zero.  This establishes the consistency of the  $ \hat F_h(t)$.
 We need the following assumptions to prove the consistency of our estimator. Let $\varphi(t)$ be the characteristic function of $X$.
\begin{enumerate}
    \item[A] There is a $p>0$ such that $\int_{-\infty}^{\infty} |t|^p |\varphi(t)| < \infty$
    \item[B] There are positive constants $d$ and $D$ such that $|\varphi(t)| \leq D e^{-d |t|}$
    \item[C]  There is a positive constant $b$ such that $\varphi(t) = 0$ when $|t| \geq b$.
\end{enumerate}
%
Next, we prove the consistency of our estimators. For this purpose first, we prove the consistency of
$\hat{\bar{F}}(t)$. Using, simple algebraic manipulation, we can see that
$$\tilde{K}(t)= 1-\int_t^ {\infty} k(u) du =1-\bar{K}(t).$$Therefore, we obtain the relationship given by
$$\hat{F}(t)=1-\hat{\bar{F}}(t).$$
Hence, in a similar way to establish the expression in (\ref{varbp}), we  have
\begin{equation}\label{varqp}
\operatorname{Var}(\hat{\bar{F}}_h(t)) = \frac{\bar{F}(t)(1 - \bar{F}(t))}{n} + \frac{2 f(t)}{n} \left( \int u \, \bar{K}(u) k(u) \, du \right) h + O\left(\frac{h}{n}\right).
\end{equation}
\doublespace
Using the  expression (\ref{varqp}), we can establishes that $\hat{\bar{F}}_h(t)$ is a consistent estimator $\bar{F}(t)$. \\
Under the assumptions $A$ to $C$ and   if  $h\to 0$ as $n\to \infty$ and $nh\to\infty$, in view of expression (\ref{est1}),  $\widehat{R}_{\alpha, \beta}(\widehat{\bar{F}},\widehat{\bar{G}})$ is a consistent estimator of $R_{\alpha, \beta}(\Bar{F},\Bar{G})$. Also in view of expression (\ref{est2}),  in accordance with (\ref{varqp}) we can show that $\widehat{R}_{\alpha,\beta}(\widehat{\bar{F}},\widehat{\bar{G}},t)$ is a consistent estimator of  $R_{\alpha, \beta}( \Bar{F},\Bar{G},t)$.

 Next,  we investigate the asymptotic distribution of $\hat{R}_{\alpha,\beta}(\hat{\bar{F}}, \hat{\bar{G}})$  using a simulation study. Figure 1 shows the empirical densities of the standardized value of $\hat{R}_{\alpha,\beta}(\hat{\bar{F}}, \hat{\bar{G}})$ generated with 100,000 samples of sizes $n=100,200,500,1000$ where $X$ and $Y$ has standard exponential distribution and
$\alpha=\beta=1$. From, Figure 1, it is evident that the limiting distribution of the standardized value of the estimator is standard normal.
\begin{figure}[h]
    \centering
    \vspace{-0.5cm}
    \includegraphics[width=0.8\linewidth]{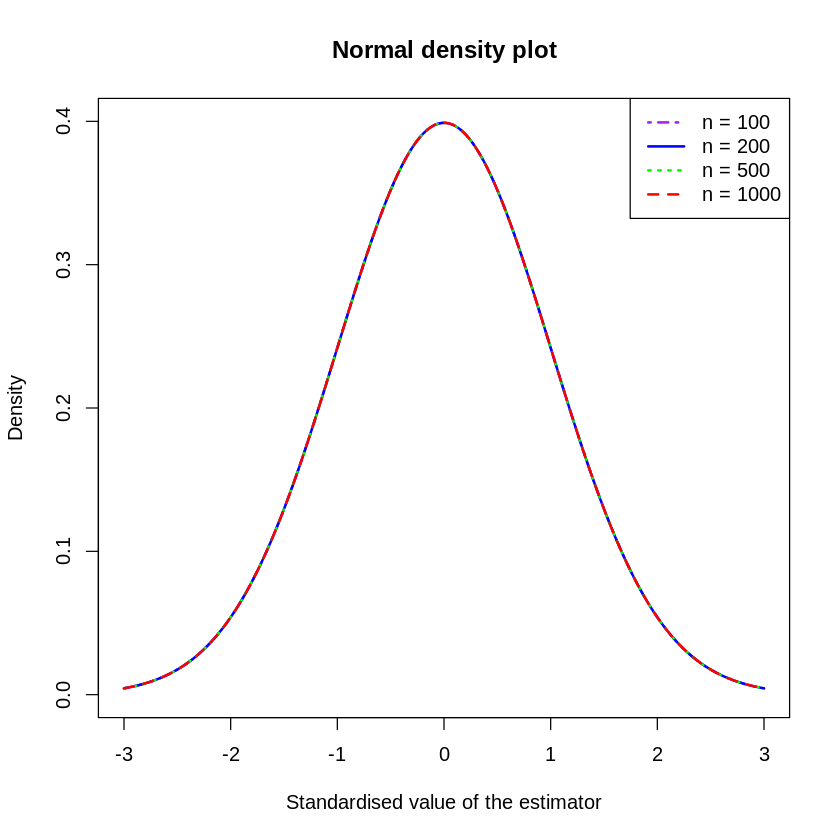}
    \vspace{-0.5cm}
    \caption{Normal density plot}
    \label{fig:enter-label}
\end{figure}
\FloatBarrier
\doublespace
\section {Simulation Studies}
This section will look into the Monte Carlo simulation studies based on the estimators $\widehat{R}_{\alpha, \beta} (\hat{\bar{F}},\hat{\bar{G}})$ and $\widehat{R}_{\alpha,\beta}(\hat{\bar{F}},\hat{\bar{G}},t)$. The simulation is conducted using R software. The experiment is repeated 10,000 times using different sample sizes, $n=10,20,30,40,50$.
For the simulation, we generate the $X$ and $Y$ using different lifetime distributions namely, exponential, Weibull, Pareto, and lognormal.  We also consider an exponential-Weibull combination where one data set is generated from exponential and the other data set from Weibull distribution.  The parameters are chosen randomly and different sample sizes are considered for various choices of $\alpha$ and $\beta$.
 Kernel survival estimator is used to find the estimates of the proposed measure. In our study, we opted the Silverman's thumb rule for selecting the bandwidth $h$ and is taken as $h = 1.06 \hat{\sigma} n^{-\frac{1}{5}}$, where $\hat{\sigma}$, is the standard deviation of the $n$ observations taken into consideration. Using (\ref{est1}) and (\ref{est2}) we find out the estimates for RCRI and DRCRI measures and eventually bias and MSE are also calculated.\\

Tables 2-5 provides the results regarding the Bias and MSE for different distributions of the estimator of RCRI measure.First, we generated two different set of  random samples from standard exponential distribution. Similarly, we took a pair of Weibull, Pareto, and lognormal random samples into consideration. From Table 2, it is observed that the exponential random samples showed a better performance than the other distributions.\\

Table 3 gives the results regarding the bias and MSE of RCRI measure when $\alpha=1 $ and $\beta=2 $. Here we can see that the Pareto distribution performed better than the other distributions.
For further evaluation, we generated random samples from  exponential and Weibull. Table 4 and Table 5 provides the results of the same. In all these cases, we can observe that  the bias and MSE decreases as $n$ increases.
\begin{table}[h!]
\centering
\caption{Bias and MSE of RCRI for different distributions, $\alpha=1$ and $\beta=1$}
\label{tab:example}
\begin{tabular}{c c c c c c c c c}
\hline
$n$ & \multicolumn{2}{c}{X$\sim $exponential(1)} & \multicolumn{2}{c}{Weibull(3,1)} & \multicolumn{2}{c}{Pareto(1,3)} & \multicolumn{2}{c}{lognormal(0,1)} \\
& \multicolumn{2}{c}{Y$\sim$ exponential(1)} & \multicolumn{2}{c}{Weibull(3,1)} & \multicolumn{2}{c}{Pareto(1,3)} & \multicolumn{2}{c}{lognormal(0,1)} \\
\cline{2-9}
 & Bias & MSE & Bias & MSE & Bias & MSE & Bias & MSE \\
\hline
10 & 0.0108 & 0.0001 & 0.0328 & 0.0010 & 0.0303 & 0.0009 & 0.0138 & 0.0404 \\
20 & 0.0072 & 0.0000 & 0.0270 & 0.0007 & 0.0249 & 0.0006 & 0.0008 & 0.0189 \\
30 & 0.0062 & 0.0000 & 0.0230 & 0.0005 & 0.0223 & 0.0005 & 0.0078 & 0.0120 \\
40 & 0.0049 & 0.0000 & 0.0209 & 0.0004 & 0.0206 & 0.0004 & 0.0116 & 0.0091 \\
50 & 0.0043 & 0.0000 & 0.0190 & 0.0003 & 0.0198 & 0.0003 & 0.0146 & 0.0071 \\
\hline
\end{tabular}
\end{table}

\FloatBarrier

\begin{table}[h!]
\centering
\caption{Bias and MSE of RCRI for different distributions, $\alpha=1$ and $\beta=2$}.
\label{tab:example}
\begin{tabular}{c c c c c c c c c}
\hline
$n$ & \multicolumn{2}{c}{X$\sim $exponential(0.1)} & \multicolumn{2}{c}{Weibull(3,1)} & \multicolumn{2}{c}{Pareto(2,3)} & \multicolumn{2}{c}{lognormal(0.5,0.5)} \\
& \multicolumn{2}{c}{Y$\sim$ exponential(0.8)} & \multicolumn{2}{c}{Weibull(3,1)} & \multicolumn{2}{c}{Pareto(2,3)} & \multicolumn{2}{c}{lognormal(0.5,0.5)} \\
\cline{2-9}
 & Bias & MSE & Bias & MSE & Bias & MSE & Bias & MSE \\
\hline
10 & 0.1792 & 0.0321 & 0.0454 & 0.0020 & 0.0153 & 0.0002 & 0.0787 & 0.0219 \\
20 & 0.1428 & 0.0204 & 0.0386 & 0.0014 & 0.0091 & 0.0000 & 0.0896 & 0.0157 \\
30 & 0.1247 & 0.0155 & 0.0336 & 0.0011 & 0.0068 & 0.0000 & 0.0922 & 0.0922 \\
40 & 0.1120 & 0.0125 & 0.0307 & 0.0009 & 0.0059 & 0.0000& 0.0928 & 0.0124 \\
50 & 0.1031 & 0.0106 & 0.0282 & 0.0007 & 0.0052 & 0.0000 & 0.0924 & 0.0116 \\
\hline
\end{tabular}
\end{table}

\FloatBarrier

\begin{table}[h!]
\centering
\caption{Bias and MSE of RCRI for different distributions, $\alpha=1$ and $\beta=1$}
\label{tab:example}
\begin{tabular}{c c c c c c c}
\hline
$n$ & \multicolumn{2}{c}{X$\sim $exponential(1)} & \multicolumn{2}{c}{exponential(3)} & \multicolumn{2}{c}{exponential(3)}  \\
& \multicolumn{2}{c}{Y$\sim$ Weibull(1,1)} & \multicolumn{2}{c}{Weibull(5,1)} & \multicolumn{2}{c}{Weibull(5,2)} \\
\cline{2-7}
 & Bias & MSE & Bias & MSE & Bias & MSE \\
\hline
10 & 0.0120 & 0.0001 & 0.0174 & 0.0003 & 0.1016 & 0.0103 \\
20 & 0.0067 & 0.0000 & 0.0138 & 0.0002 & 0.0836 & 0.0069 \\
30 & 0.0060 & 0.0000 & 0.0133 & 0.0002 & 0.0758 & 0.0057 \\
40 & 0.0056 & 0.0000 & 0.0122 & 0.0001 & 0.0698 & 0.0048 \\
50 & 0.0042 & 0.0000 & 0.0117 & 0.0001 & 0.0659 & 0.0043 \\
\hline
\end{tabular}
\end{table}

\begin{table}[h!]
\centering
\caption{Bias and MSE of RCRI for different distributions, $\alpha=1$ and $\beta=2$}
\label{tab:example}
\begin{tabular}{c c c c c c c}
\hline
$n$ & \multicolumn{2}{c}{X$\sim $exponential(1)} & \multicolumn{2}{c}{exponential(3)} & \multicolumn{2}{c}{exponential(3)}  \\
& \multicolumn{2}{c}{Y$\sim$ Weibull(1,1)} & \multicolumn{2}{c}{Weibull(5,1)} & \multicolumn{2}{c}{Weibull(5,2)} \\
\cline{2-7}
 & Bias & MSE & Bias & MSE & Bias & MSE \\
\hline
10 & 0.0182 & 0.0003 & 0.0055 & 0.0000 & 0.0221 & 0.0005 \\
20 & 0.0216 & 0.0005 & 0.0024 & 0.0000 & 0.0158 & 0.0002 \\
30 & 0.0208 & 0.0004 & 0.0020 & 0.0000 & 0.0138 & 0.0002 \\
40 & 0.0201 & 0.0004 & 0.0014 & 0.0000& 0.0120 & 0.0001 \\
50 & 0.0197 & 0.0004 & 0.0014 & 0.0000& 0.0112 & 0.0001 \\
\hline
\end{tabular}
\end{table}

\FloatBarrier

Next, we study the performance of the estimator of the DRCRI measure for different values of $t$ and $n$. Table 6 provides the results regarding the bias and MSE of DRCRI when $X \sim \text{exponential}(1)$ and $Y \sim \text{exponential}(1)$, with $\alpha = 1$ and $\beta = 1$.
The investigation also considered evaluating the bias and MSE of DRCRI for different $t$ and $n$ values when $X \sim \text{exponential}(1)$ and $Y \sim \text{Weibull}(5, 3)$, with $\alpha = 1$ and $\beta = 2$. The results are provided in Table 7.

All the results showed a decrease in the value of bias and MSE as the value of $n$ increases.

\begin{table}[h!]
\centering
\caption{Bias and MSE of DRCRI for different \( t \) and \( n \) values when $X\sim $ exponential(1) and $Y\sim$ exponential(1), $\alpha=1$ and $\beta=1$}
\label{tab:example}
\begin{tabular}{c c c c}
\hline
$t$ & $n$ & Bias & MSE \\
\hline
\multirow{5}{*}{0.5} & 10 &  0.0887 & 0.0361 \\
                      & 20 & 0.0710 & 0.0178 \\
                      & 30 & 0.0606 & 0.0118 \\
                      & 40 & 0.0525 &0.0086 \\
                      & 50 & 0.0468 & 0.0067 \\
\hline
\multirow{5}{*}{0.75}  & 10 & 0.0389 & 0.0289\\
                      & 20 & 0.0300 & 0.0132 \\
                      & 30 & 0.0243 & 0.0085 \\
                      & 40 & 0.0195 & 0.0060 \\
                      & 50 & 0.0162 & 0.0046 \\
\hline
\multirow{5}{*}{1} & 10 & 0.0015 & 0.0275 \\
                      & 20 & 0.0035 & 0.0125 \\
                      & 30 &0.0029 & 0.0081 \\
                      & 40 & 0.0014 & 0.0059 \\
                      & 50 & 0.0006 & 0.0046 \\

\hline
\end{tabular}
\end{table}

\begin{table}[h!]
\centering
\caption{Bias and MSE of DRCRI for different \( t \) and \( n \) values when $X\sim exponential(1)$ and $Y\sim$ Weibull(5,3), $\alpha=1$ and $\beta=2$}
\label{tab:example}
\begin{tabular}{c c c c}
\hline
$t$ & $n$ & Bias & MSE \\
\hline
\multirow{5}{*}{0.5} & 10 & 0.0945 & 0.0092 \\
                      & 20 & 0.0424 & 0.0018 \\
                      & 30 & 0.0256 & 0.0006 \\
                      & 40 & 0.0174 & 0.0003 \\
                      & 50 & 0.0125 & 0.0001 \\
\hline
\multirow{5}{*}{0.75}  & 10 & 0.0909 & 0.0086\\
                      & 20 & 0.0432 & 0.0019 \\
                      & 30 & 0.0279 & 0.0007 \\
                      & 40 & 0.0204 & 0.0004 \\
                      & 50 & 0.0159 & 0.0002 \\
\hline
\multirow{5}{*}{1} & 10 & 0.0849 & 0.0075 \\
                      & 20 & 0.0415 & 0.0017 \\
                      & 30 & 0.0272 & 0.0007 \\
                      & 40 & 0.0202 & 0.0004 \\
                      & 50 & 0.0161 & 0.0002 \\

\hline
\end{tabular}
\end{table}
\FloatBarrier
Next, we conduct a comparison study of the estimator of the RCRI measure and the estimators of existing measures.

\begin{table}[ht]
\centering
\caption{Comparison of RCRI estimator with the existing measures}
\label{tab:example}
\renewcommand{\arraystretch}{1.2}
\begin{tabular}{lcccccc}
\hline
\multicolumn{5}{c}{$(\lambda_1,\lambda_2)=(0.1,0.8)$} \\
\hline
Measure / $n$ & 10 & 20 & 30 & 40 & 50 & 100 \\
\hline
$\widehat{R}_{\alpha, \beta}$ & 0.0321 & 0.0204 & 0.0155 & 0.0125 & 0.0106  & 0.0060\\
$\widehat C_{oz}$
  & 0.0215 & 0.0169 & 0.0141 & 0.0132 &0.0120& 0.0085\\
$\widehat D_{k}
$      & 3.3052 & 1.6521 &1.0315 & 0.7780 &0.6450& 0.3234\\
\hline

\multicolumn{5}{c}{$(\lambda_1,\lambda_2)=(1,1)$} \\
\hline
Measure / $n$ & 10 & 20 & 30 & 40 & 50 & 100 \\
\hline
$\widehat{R}_{\alpha, \beta}$ & 0.0001 & 0.0000 & 0.0000 & 0.0000 & 0.0000 & 0.0000 \\
$\widehat C_{oz}$  & 0.0702 & 0.0338 & 0.0234 & 0.0172 & 0.0146 & 0.0075\\
$\widehat D_{k}$      & 0.0124 & 0.0028 & 0.0012 & 0.0006 & 0.0005 & 0.0001\\
\hline

\multicolumn{5}{c}{$(\lambda_1,\lambda_2)=(0.1,0.5)$} \\
\hline
Measure / $n$ & 10 & 20 & 30 & 40 & 50 & 100 \\
\hline
$\widehat{R}_{\alpha, \beta}$ & 0.0060 & 0.0014 & 0.0005 & 0.0001 & 0.0000
& 0.0000
\\
$\widehat C_{oz}$  & 0.0222 & 0.0135 & 0.0112 & 0.0093 & 0.0078 & 0.0052\\
$\widehat D_{k}$      & 3.2353 & 1.5888 & 0.9834 & 0.7456 & 0.6205 & 0.3068 \\
\hline

\multicolumn{5}{c}{$(\lambda_1,\lambda_2)=(1,5)$} \\
\hline
Measure / $n$ & 10 & 20 & 30 & 40 & 50 & 100 \\
\hline
$\widehat{R}_{\alpha, \beta}$ & 0.0042 & 0.0029 & 0.0023 & 0.0020 & 0.0018 & 0.0012\\
$\widehat C_{oz}$  & 0.0206 & 0.0133 & 0.0105 & 0.0087 & 0.0079 & 0.0053 \\
$\widehat D_{k}$      & 0.0323 & 0.0158 & 0.0098 & 0.0074 & 0.0062 & 0.0030 \\
\hline

\multicolumn{5}{c}{$(\lambda_1,\lambda_2)=(10,9)$} \\
\hline
Measure / $n$ & 10 & 20 & 30 & 40 & 50 & 100 \\
\hline
$\widehat{R}_{\alpha, \beta}$ & 0.0000 &0.0000 & 0.0000 & 0.0000 & 0.0000 & 0.0000 \\
$\widehat C_{oz}$  & 0.0519 & 0.0212 & 0.0130 & 0.0101 &0.0076 & 0.0031\\
$\widehat D_{k}$      & 0.0001 & 0.0000 &0.0000 & 0.0000 &0.0000&0.0000 \\
\hline
\end{tabular}
\end{table}
\FloatBarrier
Here  we compare the mean squared error (MSE) of the estimator for the measure $RCRI$ with the estimators of $C_{oz}$ \citet{cox2016practical} and  $D_{k}$ a novel measure developed by \citet{garg2025empirical}.

\begin{equation}
C_{oz} = \int_{0}^{\infty} {\bar F(x) g(x)}-{\bar G(x) f(x)} \, dx
\end{equation}

\begin{equation}
D_{k} = \int_{0}^{\infty} \left( \bar F(x) - \bar G(x) \right)^{2} \, dx
\end{equation}

The corresponding estimators of the above measures are given below .
\[
\widehat C_{oz} = \int_{0}^{\infty} \bar F_{n}(x) g_{n}(x)-{\bar G_{n}(x)\, f_{n}(x)} \, dx
\]
 and
 \[
\widehat D_{k} = \int_{0}^{\infty} \left( \bar F_{n}(x) - \bar G_{n}(x) \right)^{2} \, dx
\]
where $f_{n}(x)$, $g_{n}(x)$,$\bar F_{n}(x)$ and $\bar G_{n}(x)$ are defined in Section 4. Two exponential distributions with parameters $\lambda_1 $ and $\lambda_2$ are taken into consideration. To evaluate the comparative performance of the three estimators, $\widehat{R}_{\alpha, \beta}$, $\widehat C_{oz}$ and
$\widehat D_{k}$ we compute their mean squared errors (MSEs) with respect to the true values across a range of sample sizes and parameter settings. The proposed estimator $\widehat{R}_{\alpha, \beta}$ consistently exhibits the smallest MSE across all parameter combinations and sample sizes, demonstrating superior accuracy and stability relative to $\widehat C_{oz}$ and $\widehat D_{k}$.
The results are presented in Table 8.

\section {Data Analysis}
Gaia is a European space mission provides astrometry, photometry, and spectroscopy of nearly 2 billion stars in the Milky Way as well as significant samples of extra galactic and solar system objects. The third Gaia data release, Gaia DR3, contains astrometry and broad-band photometry, \citet{brown2016gaia}, \citet{vallenari2023gaia}, \citet{steen2024measuring}.\\
We considered data collected between 25 July 2014 and 28 May 2017 - during the first 34 months of the Gaia mission, which have been processed by the Gaia Data Processing and Analysis Consortium (DPAC), resulting in Gaia DR3 for data analysis.
In this study, we have taken into account the epoch photometry of the Gaia DR3 object 4111834567779557376 and the epoch photometry of the Gaia DR3 object 5090605830056251776. The epoch photometry table contains the light curve for a given object in the pass bands G, BP, and RP. The data related to the magnitude of the pass bands was taken from the Gaia DR3 archive \url{https://gea.esac.esa.int/archive/}.
The objective here is to compare the magnitude of various pass bands namely G, BP, and RP. We have estimated the results of RCRI measure for the pairs (G, BP) (G, RP) and (BP, RP). For both Gaia DR3 4111834567779557376 and Gaia DR3  5090605830056251776 a total of 150 magnitude observations were considered, comprising 50 observations each for the G, BP, and RP passbands. Tables 9 and 10 present the estimated RCRI values for the pairs (G, BP), (G, RP), and (BP, RP) corresponding to Gaia DR3 4111834567779557376 and Gaia DR3  5090605830056251776 respectively.  It can be seen that the disparity between the pairs of pass bands are consistent, while considering both the objects.
\begin{table}[h!]
\centering
\caption{RCRI of the pairs of (G, BP) (G, RP) and (BP, RP)\\
Gaia DR3 4111834567779557376,  $\alpha=1 ,\beta=1$}
\begin{tabular}{cccc}

\(\bar{F}(x)\) & \(\bar{G}(x)\) & RCRI \\
\hline
G & BP& 7.0291& \\
G & RP & 6.3994 & \\
BP & RP& 6.4007 & \\

\hline
\end{tabular}

\label{table:example}
\end{table}
\FloatBarrier
\begin{table}[h!]
\centering
\caption{RCRI of the pairs of (G, BP) (G, RP) and (BP, RP)\\
Gaia DR3 5090605830056251776,  $\alpha=1 ,\beta=1$}
\begin{tabular}{cccc}

\(\bar{F}(x)\) & \(\bar{G}(x)\) & RCRI \\
\hline
G & BP&  17.5250& \\
G & RP & 17.1916 & \\
BP & RP& 17.0297 & \\

\hline
\end{tabular}

\label{table:example}
\end{table}




 These results suggest that the magnitude distributions corresponding to the G, BP, and RP passbands exhibit broadly comparable pairwise information characteristics for the two Gaia DR3 objects considered in this study.
 \section {Application of RCRI Measure in Image Processing}
To ensure the accuracy and reliability of our results, we have taken into account six chest X-ray images from the publicly available chest X-ray 2017 dataset, contributed by \citet{kermany2018labeled}. These images include samples from healthy individuals and patients diagnosed with pneumonia. For analysis, each X-ray image was converted into a $50\times 50$ pixel grid, resulting in 2500 cells per image. The intensity of the gray level of each cell was recorded as a real number ranging from 0 (black) to 1 (white), effectively capturing the grayscale distribution of the image. The $\alpha$ and $\beta$ are take as $5$.

Using these six images, we generated all 36 possible pairwise combinations. For each pair, we are estimating the Relative Cumulative Residual Information (RCRI), which serves as a measure of discrepancy between the Gray level distributions of the images.

As illustrated in Figure 2, images (a), (b), and (c) correspond to normal chest X-rays, while images (d), (e) and (f) represent pneumonia-infected cases. The calculated RCRI values are presented in Table 11. The matrix reveals distinct clustering patterns that effectively separate the two groups. The top-left $3\times 3$ submatrix, which includes comparisons among normal images (a, b, c), shows RCRI values ranging from 0.0559 to 0.0637, indicating strong similarity within this group. Similarly, the bottom-right $3\times 3$ submatrix, corresponding to comparisons among pneumonia-infected images (d, e, f), shows moderately higher but internally consistent RCRI values between 0.1854 and 0.2533, reflecting similarity within this second group. In contrast, RCRI values for comparisons between normal and pneumonia images fall into an intermediate range (approximately 0.0858 to 0.1204), distinctly separating the two classes. These patterns support the application of RCRI as an effective statistical measure for distinguishing grayscale intensity distributions, highlighting its potential in medical image classification and diagnostic support. It should be noted that the secondary nature of the medical images restricts the extent to which definitive conclusions regarding the diagnostic capability of the proposed measure can be drawn. Future studies based on primary X-ray images acquired under standardized conditions, such as uniform image size, quality, imaging equipment, and operator protocols, could provide a more reliable basis for developing diagnostic criteria. In particular, the RCRI measure demonstrates considerable potential for distinguishing between normal and pneumonia-infected images, and its evaluation using standardized data may facilitate the establishment of a critical value or critical interval for diagnostic classification.

\begin{figure}[htbp]
  \centering
  \begin{subfigure}[b]{0.3\textwidth}
    \includegraphics[width=\linewidth,height=40mm]{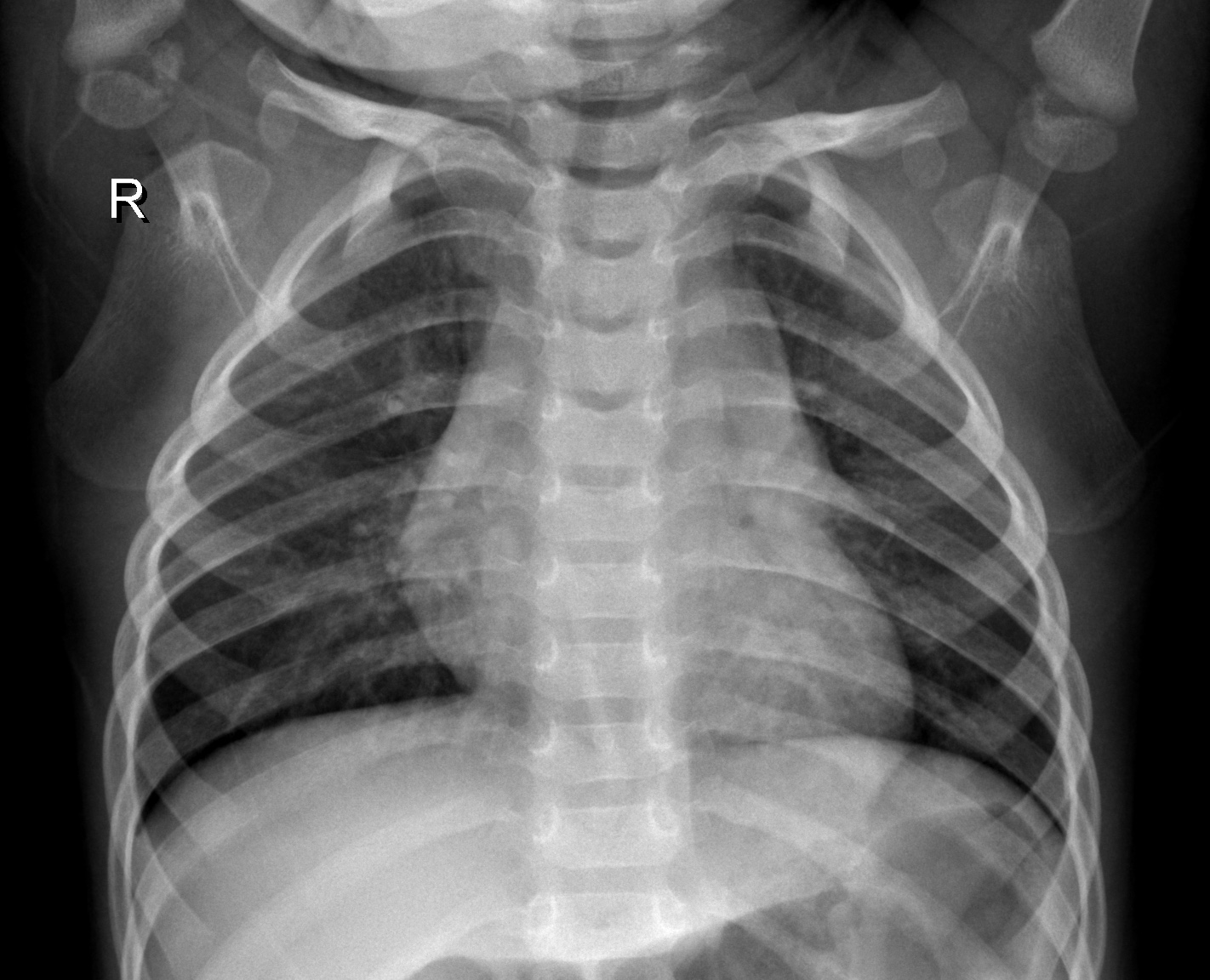}
  \caption{}
  \end{subfigure}
  \hfill
  \begin{subfigure}[b]{0.3\textwidth}
    \includegraphics[width=\linewidth,height=40mm]{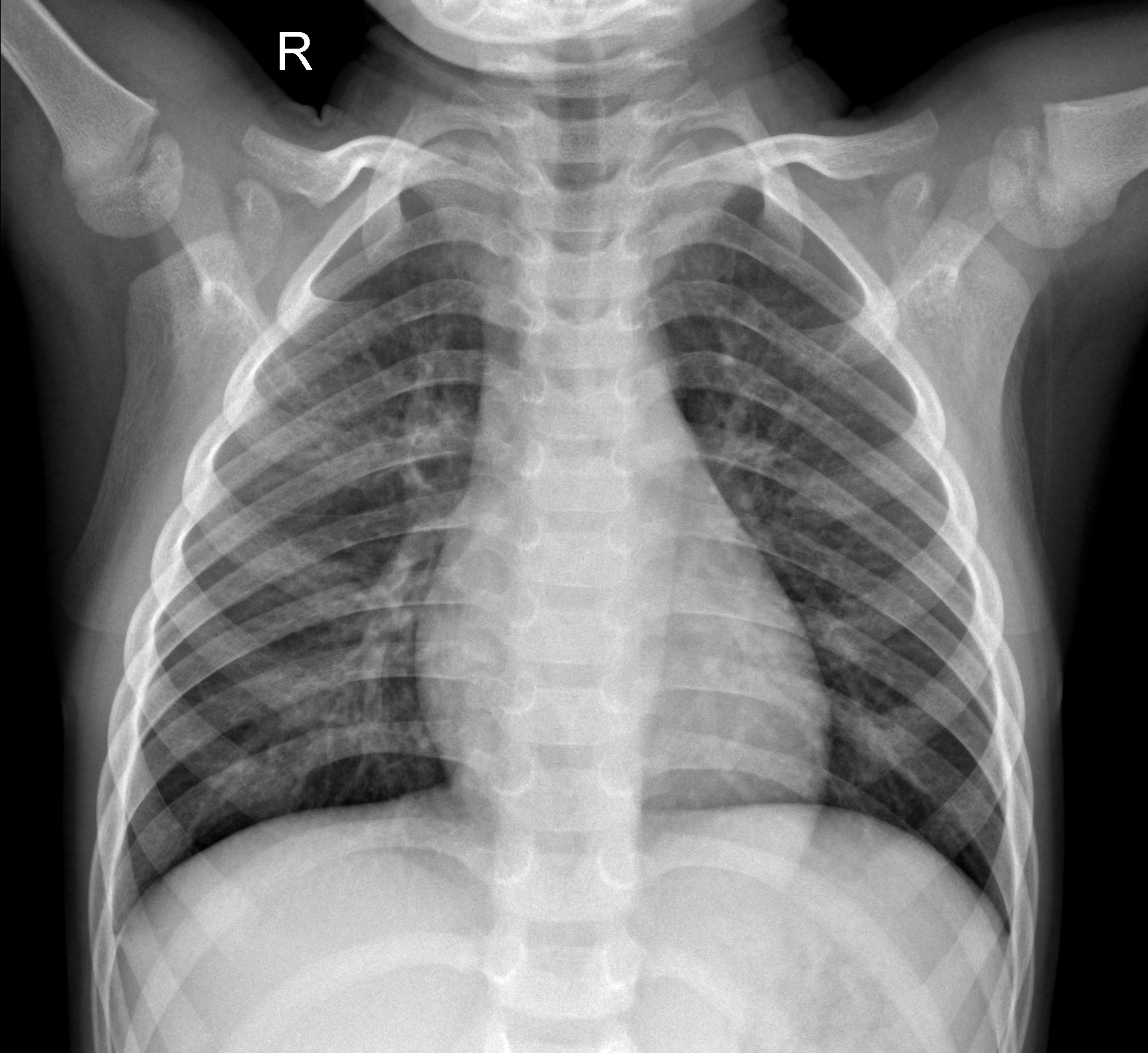}
  \caption{}
  \end{subfigure}
  \hfill
  \begin{subfigure}[b]{0.3\textwidth}
    \includegraphics[width=\linewidth,height=40mm]{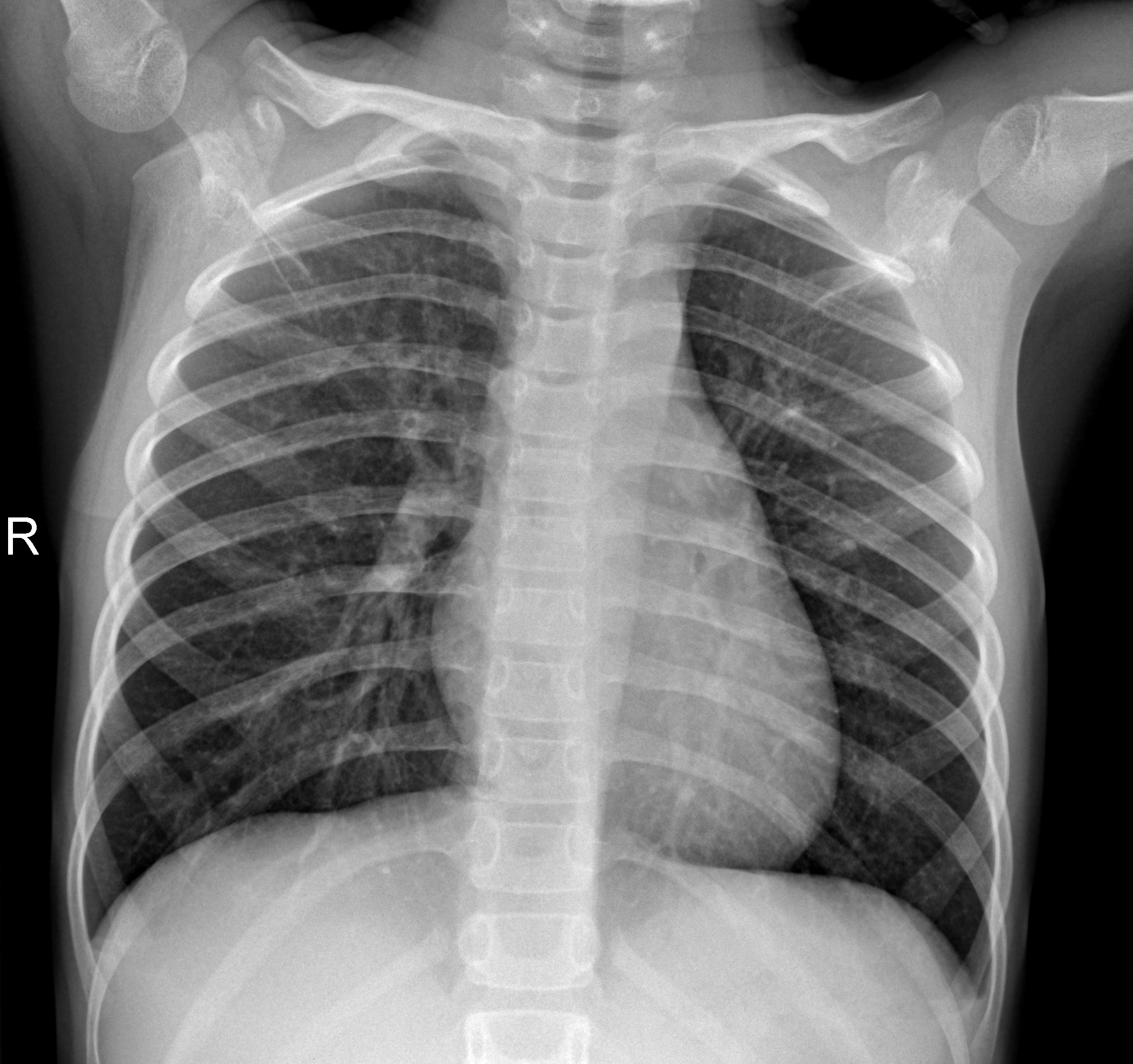}
    \caption{}
  \end{subfigure}
  \vspace{0.9cm}
  \begin{subfigure}[b]{0.3\textwidth}
    \includegraphics[width=\linewidth,height=40mm]{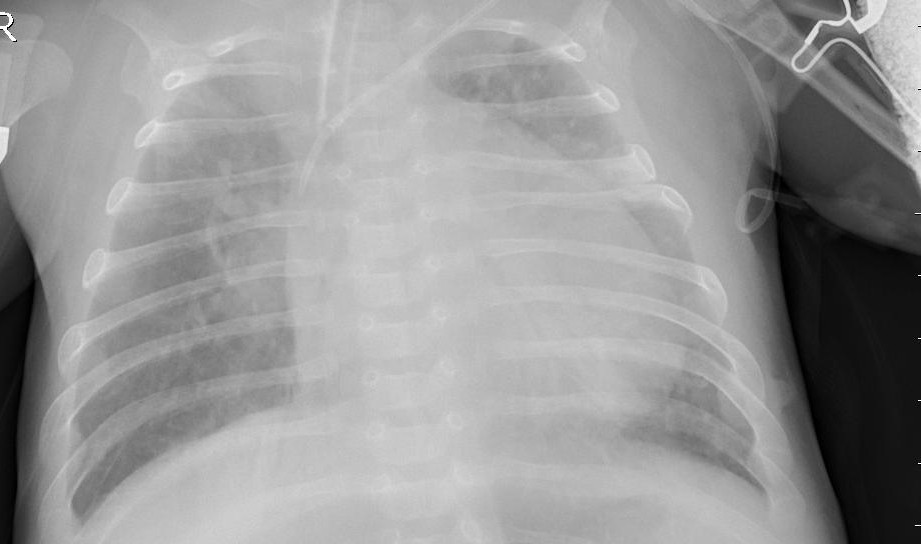}
    \caption{}
  \end{subfigure}
  \hfill
  \begin{subfigure}[b]{0.3\textwidth}
    \includegraphics[width=\linewidth,height=40mm]{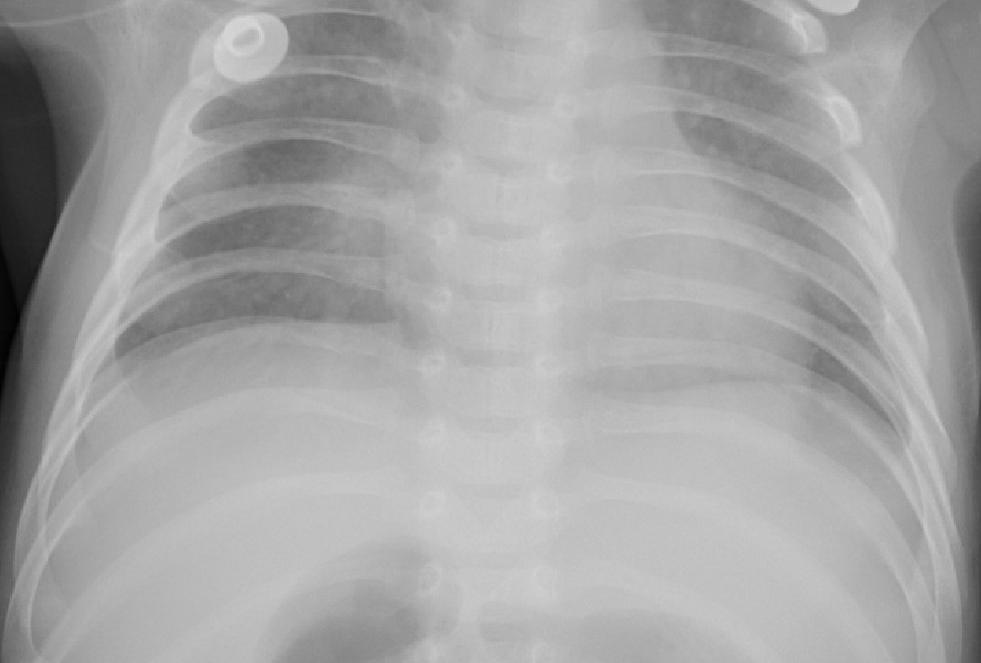}
    \caption{}
  \end{subfigure}
  \hfill
  \begin{subfigure}[b]{0.3\textwidth}
    \includegraphics[width=\linewidth,height=40mm]{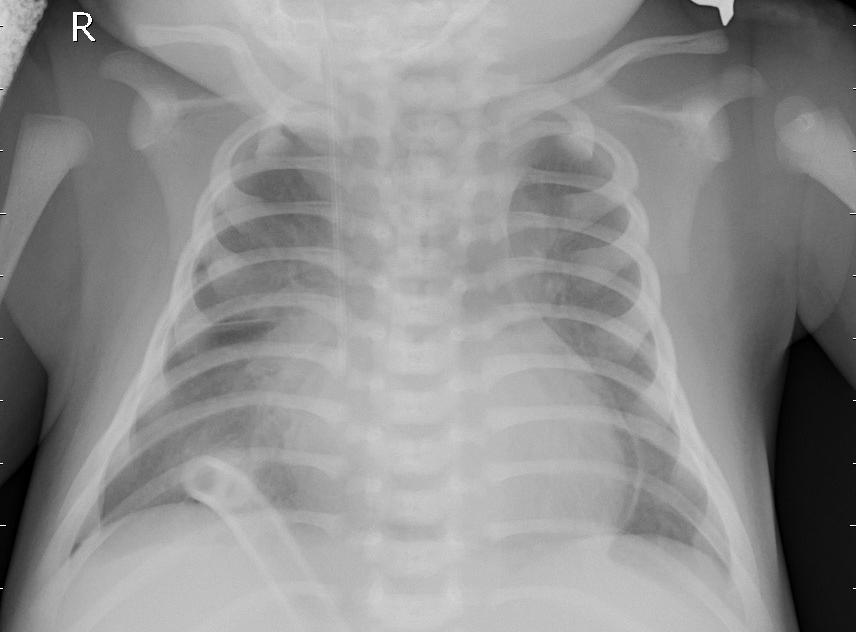}
    \caption{}
  \end{subfigure}
  \caption{Chest X-ray Images}
  \label{fig:chest_xrays}
\end{figure}
\begin{table}[h!]
\centering
\caption{Matrix of Relative Cumulative Residual Information (RCRI) for Chest X-ray Image Pairs}
\label{tab:rcri_matrix}
\begin{tabular}{c|cccccc}
     & a       & b       & c       & d       & e       & f       \\
\hline
a & 0.0512 & 0.0564 & 0.0559 & 0.0858 & 0.0983 & 0.0955 \\
b & 0.0564 & 0.0652 & 0.0637 & 0.1030 & 0.1204 & 0.1156 \\
c & 0.0559 & 0.0637 & 0.0625 & 0.0996 & 0.1159 & 0.1114 \\
d & 0.0858 & 0.1030 & 0.0996 & 0.1854 & 0.2150 & 0.1984 \\
e & 0.0983 & 0.1204 & 0.1159 & 0.2150 & 0.2533 & 0.2334 \\
f & 0.0955 & 0.1156 & 0.1114 & 0.1984 & 0.2334 & 0.2197 \\
\end{tabular}
\end{table}
\FloatBarrier
\section {Conclusion}
 In this paper, we developed the extended concept of information generating measure namely relative cumulative residual information (RCRI), and also a dynamic version of the same has been discussed (DRCRI). Several theorems and propositions based on the above measures are studied in detail. An upper bound for the RCRI measure in terms
of cumulative residual entropy generating functions is obtained using the arithmetic mean
and geometric mean inequality.  The results of the characterization of the relationship between DRCRI and the hazard rate are examined. Furthermore, a nonparametric estimator, kernel density estimator, curated for the survival function, is obtained for RCRI and DRCRI measures. The performance of both are evaluated.We evaluated the bias and MSE of the estimator using Monte Carlo simulation method. Practical applications of the RCRI measure are illustrated using the epoch photometry data collected from the third Gaia data release, Gaia DR3.  Furthermore, the image classification based on grayscale intensity distributions using the RCRI measure was implemented using chest X-ray data from the publicly available dataset contributed by \citet{kermany2018labeled}. Exploring weighted generating functions in the framework introduced by \citet{mohammadi2024weighted} remains an open problem.

\section*{Acknowledgment}
This work has utilized data from the European Space Agency (ESA) mission,
processed by the {\it Gaia} Data Processing and Analysis Consortium (DPAC).

\bibliographystyle{apalike}       
\bibliography{references}         

\end{document}